\documentclass{article}

\usepackage{arxiv}

\usepackage[utf8]{inputenc} % allow utf-8 input
\usepackage[T1]{fontenc}    % use 8-bit T1 fonts
\usepackage{hyperref}       % hyperlinks
\usepackage{url}            % simple URL typesetting
\usepackage{booktabs}       % professional-quality tables
\usepackage{amsfonts}       % blackboard math symbols
\usepackage{nicefrac}       % compact symbols for 1/2, etc.
\usepackage{microtype}      % microtypography
\usepackage{cleveref}       % smart cross-referencing
\usepackage{lipsum}         % Can be removed after putting your text content
\usepackage{graphicx}
\usepackage[numbers]{natbib}
\usepackage{doi}
\usepackage{color}

\usepackage{float}

\newcommand{\C}[1]{{\textcolor{black}{#1}}}

\title{Evidence of equilibrium dynamics in human social networks evolving in time}

\newif\ifuniqueAffiliation
\ifuniqueAffiliation % Standard variant of author block
\author{ \href{https://orcid.org/0000-0000-0000-0000}{\includegraphics[scale=0.06]{orcid.pdf}\hspace{1mm}Miguel A. González-Casado}\thanks{Use footnote for providing further
		information about author (webpage, alternative
		address)---\emph{not} for acknowledging funding agencies.} \\
	Department of Computer Science\\
	Cranberry-Lemon University\\
	Pittsburgh, PA 15213 \\
	\texttt{hippo@cs.cranberry-lemon.edu} \\
	\And
	\href{https://orcid.org/0000-0000-0000-0000}{\includegraphics[scale=0.06]{orcid.pdf}\hspace{1mm}Elias D.~Striatum} \\
	Department of Electrical Engineering\\
	Mount-Sheikh University\\
	Santa Narimana, Levand \\
	\texttt{stariate@ee.mount-sheikh.edu} \\
}
\else
\usepackage{authblk}

\author[a,1]{Miguel A. González-Casado}
\author[b,c]{Andreia Sofia Teixeira}
\author[a,d]{Angel Sánchez}

\affil[a]{Grupo Interdisciplinar de Sistemas Complejos (GISC), Departamento de Matemáticas, Universidad Carlos III de Madrid, 28911 Legan\'es, Spain}
\affil[b]{Network Science Institute, Northeastern University London, London, E1W 1LP, United Kingdom}
\affil[c]{LASIGE, Departamento de Inform\'atica, Faculdade de Ci\^encias, Universidade de Lisboa, Portugal}
\affil[d]{Instituto de Biocomputaci\'on y F\'isica de Sistemas Complejos (BIFI), Universidad de Zaragoza, 50018 Zaragoza, Spain}
\affil[1]{To whom correspondence should be addressed. E-mail: miguelangel.gonzalezc@outlook.es}

\hypersetup{
pdftitle={Evidence of equilibrium dynamics in human social networks evolving in time},
pdfsubject={TBD},
pdfauthor={Miguel A. González-Casado, Andreia Sofia Teixeira, Angel Sánchez},
pdfkeywords={Personal Relationships , Social Network Analysis , Longitudinal Data , Equilibrium , Network Dynamics},
}

\begin{document}
\maketitle

\begin{abstract}
How do networks of relationships evolve over time? We analyse a dataset tracking the social interactions of 900 individuals over four years. Despite continuous shifts in individual relationships, the macroscopic structural properties of the network remain stable, fluctuating within predictable bounds. We connect this stability to the concept of equilibrium in statistical physics. Specifically, we demonstrate that the probabilities governing network dynamics are stationary over time, and key features like degree, edge, and triangle abundances align with theoretical predictions from equilibrium dynamics. Moreover, the dynamics satisfies the detailed balance condition. Remarkably, equilibrium persists despite constant turnover as people join, leave, and change connections. This suggests that equilibrium arises not from specific individuals but from the balancing act of human needs, cognitive limits, and social pressures. Practically, this equilibrium simplifies data collection, supports methods relying on single network snapshots (like Exponential Random Graph Models), and aids in designing interventions for social challenges. Theoretically, it offers new insights into collective human behaviour, revealing how emergent properties of complex social systems can be captured by simple mathematical models.
\end{abstract}

\keywords{Personal Relationships \and Social Network Analysis \and Longitudinal Data \and Equilibrium \and Network Dynamics}

\section{Introduction}
Human social behaviour is driven by a complex interplay of cognitive, emotional, and social factors, which together shape the structure and dynamics of social networks. Prior research has shown that individuals tend to maintain a finite number of social bonds constrained by their cognitive capacity \cite{Powell2012}, and how these social bonds are structured is explained by the way humans allocate their cognitive resources to create functional relationships able to cover their social necessities \cite{Tamarit2018,Tamarit2022}. From a dynamical point of view, individuals continually adjust their social ties in response to emotional, environmental, and cognitive changes. Our work goes beyond the temporal evolution of personal social relationships, focusing on whether these complex tendencies give rise to stable trends in the emergent social structure. 

In the context of social networks \cite{Wasserman1994}, personal relationship networks are defined by nodes representing people and links representing personal relationships, such as friendships or enmities. Previous research has studied the dynamics of these personal networks, specially using data collected in closed environment through surveys. Examples include the Newcomb's Fraternity data \cite{Nordlie1958,Newcomb1961,Nakao1993,Doreian1996,Doreian2001}, and Sampson's data \cite{Sampson1968}. Other studies have collected data in school environments \cite{Katz1959,Hallinan1976,Hallinan1977,Hallinan1978,Baerveldt2014,Kucharski2018}, the National Longitudinal Study of Adolescent to Adult Health (Add Health) \cite{Jeon2015,Harris2019} being specially relevant. On the other hand, networks of personal relationships have been inferred from phone calls \cite{Saramaki2014}, email exchanges \cite{Kossinets2006} or face-to-face interaction data \cite{Gelardi2021,Schaefer2010}. When longitudinal data is lacking, network dynamics can be analyzed indirectly using cross-sectional data. The basic idea behind this approach is that the dynamical mechanisms driving the temporal evolution of a network leave a fingerprint in the structure observed at single points in time. This is the key idea behind Exponential Random Graph Models \cite{Lusher2013} and related methods \cite{Cimini2019,Belaza2017,Belaza2019}. 

This entire corpus of research has led to the conclusion that three main factors drive social network dynamics \cite{Rivera2010}: (1) attributes of people (people have a number of individual attributes that are not encoded a priori in the network structure in any way and affect the way the network evolves over time, such as shared interests or social traits); (2) endogenous mechanisms (the current structure of the network constrains and modifies the future one, i.e., certain links that appear or disappear by the simple presence of other links in the network), and (3) contextual factors and disruptive events (every network is affected by the environment in which it is embedded, like the institutional setting). These factors operate locally in the network through what we call mechanisms, rules or processes that determine how the structural or functional properties of a network change over time. Some well known mechanisms are: (1) homophily and influence \cite{Rivera2010}, that drive nodes with similar traits to be positively connected; (2) reciprocity \cite{Rivera2010}, that favours the creation of bidirectional links; (3) transitivity and closure \cite{Rivera2010}, that promotes the closing of triangles and the formation of clusters; (4) differential popularity \cite{Yap2015}, that results in heterogeneous degree distributions shaped by individual traits, levels of activity or sociability, visibility, or preferential attachment; (5) balance \cite{Heider1946}, that promotes the creation of triangles with an even number of negative links; or (6) resource allocation \cite{Tamarit2018}, by which people maintain a finite number of bonds constrained by their resources availability, explaining the volatility of negative links, link removal to free up resources that can be reinvested in other links, or the prevention of the formation of new relationships due to a lack of resources, even if all the conditions for link formation are ideal. This last mechanism will be specially relevant to the appearance of the equilibrium in the network. 

Understanding the dynamics of personal relationships is difficult because these mechanisms operate simultaneously, but also because their behaviours are often coupled, giving rise to effects impossible to predict by analysing them separately. Examples of this include the formation of hierarchies and other combinations such as the ones studied in \cite{block2015reciprocity}, \cite{Schaefer2010},\cite{facchetti2011computing},\cite{doreian2014testing}, \cite{leskovec2010signed} or \cite{Yap2015}. It would thus seem that the combination of this myriad of behaviours would make the evolution of the system unpredictable. However, in this paper we show how, despite all this complexity, the temporal evolution of the network can be described by simple mathematical models and, more importantly, its behaviour can be predicted with very high accuracy. Specifically, we provide evidence that relationship networks exhibit a global behaviour akin to equilibrium dynamics. This means that the macroscopic, average properties of the network -- such as the distribution of personal contacts, the prevalence of patterns like edges, triangles, or larger structures, and aggregate metrics like network density -- remain constant, fluctuating around stable values, even as individual links continuously evolve at a microscopic level. 
Interestingly, there are apparent trends of stability in other aspects of human social behaviour, which suggest a natural tendency toward the formation of stable patterns and self-reinforcing dynamics. Thus, Hobbs and Burke show how social connections recover after the death of a friend \cite{Hobbs2017}, while Alessandreti et al. found that certain mobility patterns are conserved over time  \cite{Alessandretti2018}, linking these patterns to individuals’ social ties. Other studies, such as \cite{Crectu2022,Iniguez2023,Avalle2024,martin2018recurrent} stress the stability and robustness of online social behaviours. We will circle back to this research in the discussion section.

Despite its importance, to the best of our knowledge, no prior research has explicitly focused on determining whether social networks are indeed in equilibrium. \C{Our work fits within the broader context of complex systems methods applied to social dynamics, specially from the lens of statistical mechanics \cite{castellano2009statistical,axelrod1981evolution,pastor2001epidemic}.} In this context, we present an example of an empirical social network that is in rigorous equilibrium, given by a rich dataset collected over four years. While this is a single case, we will argue in the discussion that the emergence of equilibrium in social networks can be a very general phenomenon. The importance of this result cannot be overstated. 
%The idea that human social networks may be in equilibrium is relevant from the perspective of social behaviour because it suggests that, despite the complexity of individual interactions and the interplay between cognitive, emotional, and environmental factors, the overall structure of these networks remains stable over time. This stability points to a fundamental predictability in social systems, implying that, like physical systems, human networks follow certain rules, and their evolution is highly constrained. We will discuss how these constraints may arise from cognitive and material resource management, with individuals regulating their social ties to avoid overwhelming their social capacities. Besides, 
If a social network is in equilibrium, its properties can be predicted more accurately over time, facilitating intervention studies. Interventions can be introduced and compared against the equilibrium network, which serves as a baseline for analysis, helping to strengthen social cohesion or address social isolation. The possibility that this equilibrium behaviour may be general also suggests that certain social behaviours and organization principles transcend cultural, situational or individual differences, connecting individual interactions and broader social structures. From a modelling perspective, equilibrium assumptions enable the use of simpler, more tractable mathematical models, such as statistical mechanics tools, developed for physical systems. Conclusions about the mechanisms driving network dynamics, structural trends, and other insights apply not only to a single observed point in time but to the entire unobserved evolution of the network after the transient formation period. This is specially relevant for methods that employ cross-sectional data to investigate network dynamics, like Exponential Random Graph Models. In fact, these approaches rely on an implicit and often overlooked assumption: the network must be in or near equilibrium, as discussed in \cite{Snijders2010}. In simple terms, when using cross-sectional data, the researcher assumes that every observation is statistically equivalent to any other, a condition that is fulfilled if the network is in equilibrium.

\begin{figure}[t!]
\centering
\includegraphics[width=\linewidth]{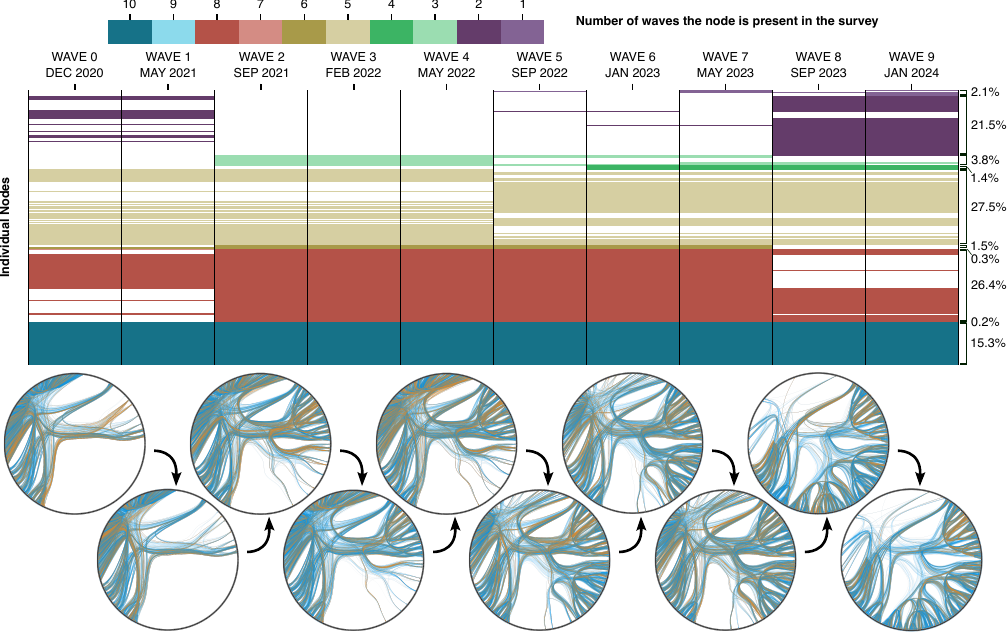}\label{fig1}
\caption{\footnotesize Representation of the network composition over the course of the 10 waves. The table in the upper part shows the presence of individuals across the waves. Each row represents a person, and each column represents a wave. A coloured cell indicates the person's presence in that specific wave. The cell colour reflects the total number of waves that person participated in. Thus, all individuals present in \textit{n} waves will have \textit{n} cells coloured in the same colour, with a different colour for each \textit{n}. Colours are not evenly distributed because each individual is likely present in all waves of the same academic year (dropouts or new enrolments in the middle of the year are rare). On the right side, the percentages of people present in each number \textit{n} of waves are displayed. For instance, 15.3\% of individuals are present in 10 waves, 0.2\% in 9 waves, etc. Below the table, a visual representation of the evolution of the network of positive and negative relationships is shown. Individuals are represented as nodes arranged in a circular layout, with blue links for positive relationships and orange links for negative ones. To illustrate the high turnover in network composition, all individuals are depicted in every wave, even if they have not yet joined or have already left the network.} 
\end{figure}

\section{Results}

In this work, we analyse a dataset we have collected between 2020 and 2024 that contains the temporal evolution of the network of personal relationships among 888 people belonging to the Blas de Otero High School in Madrid (see the Materials and Methods section for details on the data collection, data composition and data curation). The reported relationships were coded as +2 (very good), +1 (good), -1 (bad), and -2 (very bad). Almost every student provided this list, resulting in the extraction of a weighted directed network for the entire high school, which we will refer to as snapshot or wave.  We repeated this process every 16-20 weeks, and with this data we reconstructed the network at different points in time, collecting 10 snapshots of the network over 4 years. With respect to previous literature, we have introduced some improvements in the data collection process. Respondents are allowed to report an unlimited number of relationships, resulting in richer and more heterogeneous data. We also increased the sample size by one order of magnitude, with a larger number of snapshots, allowing for the detailed study of the dynamics in the long term while maintaining a certain level of granularity. Relationships are self-reported, and they are directed and weighted, between -2 and 2, allowing the consideration of negative relationships. Preliminary analyses of earlier, shorter versions of this dataset have been reported elsewhere \cite{Escribano2021,Escribano2023}. \C{It is worth noting that by coding relationships into these four categories (\(-2\), \(-1\), \(+1\), and \(+2\)), we inevitably lose many of the details that characterize social interactions. Real-world relationships exist on a spectrum, with layers of emotional, social, and contextual subtleties that cannot be completely captured in these discrete categories. Moreover, respondents may not report their relationships with the exact granularity imposed by our discretization, as these relationships are inherently multifactorial and span multiple dimensions of variability. Thus, the resulting network we analyse in this study is not a direct reflection of the complexity of real personal relationships, but rather a simplified representation. Nevertheless, even with this simplification, we argue that identifying an equilibrium in the network serves as a meaningful proxy for the underlying dynamics of real social interactions. The equilibrium we detect in this reduced representation suggests that the macroscopic patterns of stability likely reflect deeper, robust social structures. Importantly, this simplification enables the mathematical rigour of our analysis; if equilibrium behaviour is clearly observable in the simplified model, it implies a genuine stability in the underlying network of relationships, even if the full complexity of these relationships is not entirely captured.}

In total, we surveyed 888 people, but not all of them were present in all waves because the composition of the network changes as time goes by. We surveyed people belonging to a high school over the course of four academic years. Every academic year, new people enter the high school to take the first course, and at the end of the school year, almost all the people belonging to the last course leave the high school (a minor portion of them needs to repeat the course if they fail their subjects). Thus, every academic year, some nodes appear in the network, and others disappear. In Figure \ref{fig1} we depict this turnover in a visual manner. Interestingly, only 15\% of the people are present both in the first and the last waves. This will become a very relevant fact when we show the network is in equilibrium, because this equilibrium will not result from the same group of people interacting over the course of four years, but rather an internal property of the network dynamics independent of the network composition. Due to this turnover, every snapshot of the network contains about 500 active nodes out of the 888. \C{Let us define some notation for the discussions that follow. Throughout the paper, we will distinguish between a `link' and an `edge'. Although these two concepts are often treated interchangeably, here we differentiate them. A link (or tie) refers to the directional connection from one node to another, whereas an edge represents the pair of links connecting two nodes (one in each direction). We provide a more detailed explanation of this distinction later in the manuscript.}

\subsection*{Stationarity of the Transition Matrices}
In a nutshell, the concept of dynamical equilibrium in a physical system implies that the macroscopic, average properties of the system remain constant, fluctuating around a stable value, while microscopic dynamics are actively changing. For instance, in a gas, the microscopic components (the particles) are constantly moving at different velocities, colliding, vibrating, etc. However, if the gas is in equilibrium, macroscopic properties such as temperature, pressure, or volume remain constant. 

Drawing an analogy to a social system represented as a network, it is essential to define both the macroscopic properties and the microscopic dynamics. In a social network evolving over time, links appear, disappear, or change in nature. From the perspective of a node, an outgoing or incoming +1 link can become a +2, a -1, a -2 or disappear. We refer to an absent link as a 0 link. This constitutes the microscopic dynamics of the system: the evolution of individual ties/links. From these microscopic components, we can build macroscopic properties of the network. In the network, we define an edge as the connection between two nodes that contains two links, one from the first node to the second and one from the second node to the first. Since we have 5 types of links, -2, -1, 0, +1, +2, there are 25 possible edge types. Notice the distinction between link/tie and edge we introduce in our nomenclature. Although it is usual to treat both as interchangeable, we keep this distinction throughout the paper. Our first macroscopic property is the distribution of different edge types: how many +2+2, +1+2, +1+1, etc., edges exist in the network. Notice that a +2+1 edge is not equivalent to a +1+2, specially from a dynamical perspective. Although both edges can evolve towards a +1-1 edge, in the +2+1 case the +2 needs to become a +1 and the +1 a -1, and in the +1+2 case, the +1 remains unchanged and the +2 becomes a -1. Therefore, we keep this distinction in all computations. A second macroscopic property is the in-degree and out-degree distributions of the nodes. For each type of link (+2, +1, -1, and -2), we count how many nodes have 0 incoming +2 links, 1 incoming +2 links, etc. This process is repeated for outgoing links and every type of link, generating eight degree distributions. A third macroscopic property can be the distribution of different triangle types that can form among three nodes. Other macroscopic properties include structural characteristics of the network, such as clustering, average shortest path length, assortativity, centrality metrics, etc. 

In this paper, we focus on the assessment of the degree, edge, and triangle distributions since we want to be able to define transition matrices to predict the expected equilibrium states of these macroscopic properties from the dynamics observed, granting the conclusions apply to all macroscopic properties. \C{To assess whether our conclusions extend to more complex structures within the network, we have included an exemplary analysis of betweenness centrality and the details of this supplementary analysis are provided in the Supplementary Material.} Before going into the details, it is important to comment on the limitations of our approach from the statistical mechanics point of view. The main challenge is the absence of a continuous concept of time in our analysis. By relying on snapshots taken every 20 weeks, we lose information about the transitions that occur between these intervals. Hence, there is an implicit assumption that no multiple transitions between edge states have occurred within the period between snapshots, minimizing the risk of hidden dynamics. It is clear that this assumption may not hold in all cases: in systems like a gas, for example, relevant changes happen on much shorter timescales, making such large gaps between observations inappropriate. \C{This point is important because there is a close relationship between the timescale of the network dynamics, the measurement intervals, and our ability to detect equilibrium behaviour. In networks with slower dynamics, transient fluctuations may persist longer and result in more frequent apparent violations of the equilibrium, as the system might not have enough time to relax back to equilibrium between observations. Conversely, if the network dynamics are too fast, although the equilibrium state may be accurately detected, the transition matrices might not fully capture the system's internal dynamics because multiple transitions on the same link could occur between snapshots, distorting the measured change probabilities.} However, in the context of social networks, the underlying dynamics evolve more slowly, and we believe that 20-week intervals provide a sufficient resolution. On the other hand, the use of statistical mechanical techniques in this study necessarily involves certain approximations to simplify the analysis of complex social systems. We aim at striking a balance between revealing meaningful sociological patterns and maintaining the rigour without becoming overly entangled in these mathematical details. 

\C{Let us introduce the concept of transition matrix using the edges as our macroscopic property. We define the edges transition matrix $m^K$ as the matrix whose $ij$ element represents $P(j|i)$, i. e., the conditional probability of an edge of going to the $j$ state in one snapshot provided it started in the $i$ state in the previous snapshot (see Materials and Methods for the construction of this matrix). The letter $K$ labels the transition, such that $m^A$ is the transition between snapshots 0 and 1, $m^B$ is the transition between snapshots 1 and 2, etc. Additionally, we define the edge state distribution $\pi_t$ as a column vector in which each element $i$ is the density of edges of type $i$ in the snapshot $t$. With these two objects, it is direct to see that: }
\begin{equation}\label{eq1}
    \pi_{t+1} = (m^K)^T \pi_t 
\end{equation}
In other words, if we take the distribution of edges in one snapshot and multiply this distribution by the transition matrix, we obtain the distribution of edges in the next snapshot. Since we have 10 snapshots, our data allows us to construct nine transition matrices between consecutive snapshots. The first relevant question is whether these nine transition matrices are statistically equivalent. If they were, it would indicate that the dynamics are stationary, meaning the transition probabilities between edge states remain constant over time. In the Materials and Methods section we explain how to perform such a comparison.

\begin{figure}[t!]
\centering
\includegraphics[width=\linewidth]{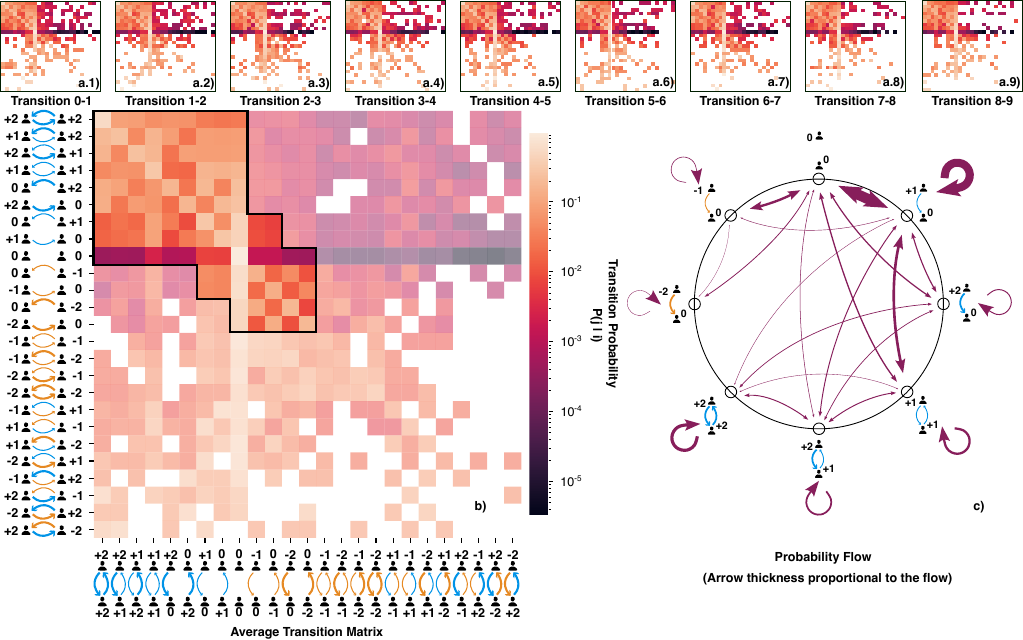}
\caption{\footnotesize Representation of the main dynamics of the network at the level of edges. In panels $a.1)-a.9)$, we depict the nine individual edge transition matrices, where each element $ij$ represents $P(j|i)$. These matrices share axes and colour bar with panel $b)$, which contains the average transition matrix. In panel $b)$, we highlight a specific part of the average transition matrix. The highlighted transitions are those with more than 50 occurrences across the 10 waves. This threshold is arbitrary, chosen for visualization purposes, but the entire matrix is included in all computations. In panel $c)$, we present a diagram illustrating the dynamics highlighted in panel $b)$. Specifically, we depict the selected edge types and use arrows to represent the probability flow between edge states, defined as $P(i)P(j|i)$. The arrow thickness is proportional to the probability flow. The self-loop of the $+0+0$ edge is not depicted because it is disproportionately large due to the network's low density. In this diagram, we have merged $+2+1$ and $+1+2$ links for simplicity of representation, although they are treated separately in computations; the same applies to $+2+0$ vs. $+0+2$, $+1+0$ vs. $+0+1$, etc. \C{In a detailed representation, there should be two arrows connecting each pair of edges to represent the probability flow in both directions. However, as we will show, the detailed balance condition is fulfilled, which implies that the two flows are nearly symmetrical. Since our intention is to present a schematic illustration that simplifies the dynamic process for clarity and ease of visualization, we use a single bidirectional arrow.}}
\label{fig2}
\end{figure}

Our findings show that the nine transition matrices are statistically equivalent (Fig. \ref{fig2}). \C{We find that for transitions 1–9, the proportions of transitions with a p-value below our significance level (see Materials and Methods) are, respectively, 0.0016, 0.0112, 0.0128, 0.0224, 0.0176, 0.0096, 0.0096, 0.0096, and 0.0144. This indicates that, although some entries in the transition matrices deviate more than expected by chance, these deviations represent only a minimal fraction of the total transitions and can be attributed to normal fluctuations. All p-values and z-scores for the statistical comparisons are provided in the paper’s repository.} Hence, we conclude that the stochastic process driving the network's evolution is stationary, with constant probabilities governing the edge changes. Figure \ref{fig2} illustrates this in panels $a.1)-a.9)$ with the individual transition matrices, and in panel $b)$ with the average transition matrix (see Materials and Methods for the construction of this average matrix), highlighting the similarity between them.

Given these results, it is natural to question whether the network is changing significantly. One might doubt whether the observed stability in dynamics is due to the network barely changing, with most edges remaining constant. For this reason, we included panel $c)$ in Figure \ref{fig2} to illustrate the main dynamics within the network. While some edges are indeed stable, such as the $+2+2$ edge, which is the most stable, there are still significant dynamics within these edges. Approximately half of the probability flow from this edge transitions to other states, showing that only a little over half of these edges persist from one wave to the next. For all other edges, the probability of transitioning to a different state is greater than that of remaining unchanged. This indicates that the dynamics are quite active, with a considerable turnover in edge states.

\begin{figure}[t!]
\centering
\includegraphics[width=0.99\linewidth]{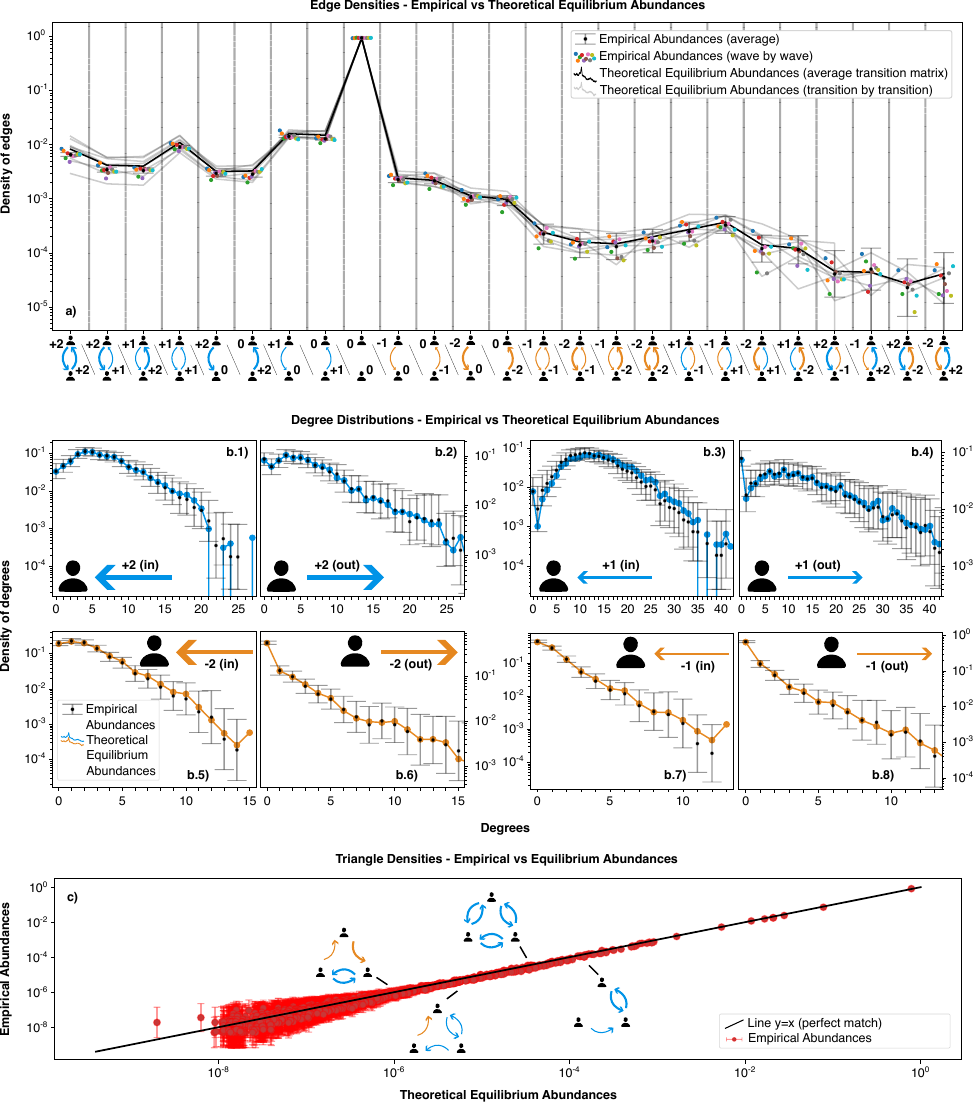}
\caption{\footnotesize Comparison between the empirical distributions of the different macroscopic properties and theoretical equilibrium distributions computed using the transition matrices. In panel $a)$ it is depicted this comparison for the edges abundances, both the averages and wave by wave, while in the rest of panels we depict only averages. In panels $b.1)-b.8)$ it is depicted this comparison for the eight different degree distributions. \C{In panel $c)$ we depict the comparison for the triangle abundances. Since there are 1924 unique triangles spanning eight orders of magnitude, we have chosen a slightly different representation for the comparison. For each triangle motif, we plot the theoretical equilibrium abundances on the x-axis and the empirical abundances on the y-axis. Thus, each point corresponds to a theoretical–empirical pair for each motif, and the closer a point lies to the y = x line, the better the match between the two. For visualization purposes, we have also included four exemplary motifs corresponding to some of the depicted points, chosen arbitrarily. In all cases, the error-bars correspond to the 95\% confidence intervals.}}
\label{fig3}
\end{figure}

This strategy can also be applied to other macroscopic properties. For instance, at the level of triangles -- similar to edges -- we can identify all unique triangle states and construct a transition matrix for these states. Given that there are five different types of directed links, the number of combinations grows exponentially with the size of the macroscopic structures analysed. In the case of triangles, we identify 1924 unique triangles (the number we observe, not the theoretical maximum). Thus, we can construct a 1924x1924 transition matrix for triangles and perform a similar test. Similarly, we can apply this method to degrees. For each node, we have eight types of degrees (in and out degrees for +2, +1, -1, and -2 links). For example, for the +2 in-degree, we examine how many nodes with an initial +2 in-degree transition to a different +2 in-degree in the next wave. This allows us to construct a transition matrix for +2 in-degrees. Although the matrix size varies depending on the specific degree analysed, the strategy remains consistent. We stress that, in all cases, we find that the transition matrices governing the evolution of these macroscopic properties are stationary, indicating that the probabilities driving the network's evolution are constant over time. \C{In all three cases (edges, triangles, and degrees), we provide the abundances, transition matrices, z-scores, and p-values in the project's repository as supplementary material.}

\begin{figure}[t!]
\centering
\includegraphics[width=0.9\linewidth]{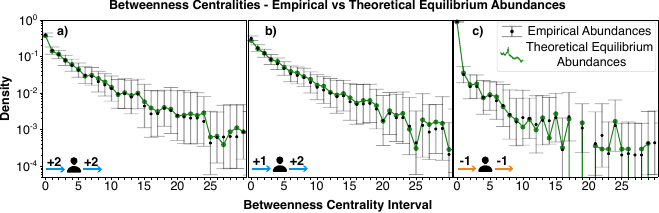}
\caption{\footnotesize \C{Comparison between the empirical distributions of the three types of the betweenness centrality considered and theoretical equilibrium distributions computed using the transition matrices. In panel $a)$ it is depicted this comparison for the centrality measured considering only +2 links. In panel $b)$ it is depicted this comparison for the centrality measured considering +2 and +1 links, treating them as equivalent. In panel $c)$ it is depicted this comparison for the centrality measured considering only -1 links. The error-bars correspond to the 95\% confidence intervals.}}
\label{fig4}
\end{figure}

\subsection*{Equilibrium State}

The stationarity of the transition matrices does not necessarily imply that the network is in equilibrium; it may not even have an equilibrium state under these dynamics. As we will see, this is not our case.

Starting from equation \ref{eq1}, focusing on edges, we can determine whether the system has a stationary state under the dynamics governed by transition matrix calculated. If it exists, we can calculate and compare it with the actual state of the network. In equation \ref{eq1}, we obtain the edge abundances in the next wave by multiplying the transpose of the transition matrix by the edge abundances in the current wave. When the system reaches the stationary state, these edge abundances become invariant under the transpose of the transition matrix. In other words, to find the stationary state associated to a transition matrix $m$ governing the dynamics, we need to solve the following equation:

\begin{equation}\label{eq2}
    \tilde{\pi} = m^T \tilde{\pi}
\end{equation}

Here, $\tilde{\pi}$ represents the stationary edge abundances, i.e., the edge abundances expected in the stationary state. Solving equation \ref{eq2} is equivalent to finding the eigenvector of the transpose of the transition matrix associated with the eigenvalue 1. A comparable procedure can be applied to obtain the stationary abundances of triangles and the eight different degree distributions. Once we obtain these theoretical abundances of edges, triangles, and degrees expected in the stationary state under the system's current dynamics, we can compare them with the current abundances to see where the system stands with respect to the stationary state. This comparison is depicted in Figure \ref{fig3}.

It is significant to note that in all cases, we find that the theoretical stationary states coincide with the observed empirical abundances, which indicates that the system has reached the stationary state. Nonetheless, to say that the system has reached the stationary state is not equivalent to say that the system has reached the equilibrium. To rigourously claim that the system is in dynamical equilibrium in the statistical mechanical sense, we need to check that the Detailed Balance Condition is fulfilled. The Detailed Balance Condition ensures that a system has reached equilibrium and is expressed as \cite{VanKampen1992}:

\begin{equation}\label{eq3}
    P(i)P(j|i) = P(j)P(i|j)
\end{equation}

Where $P(i)$ is the probability of the system being in state $i$, and $P(j|i)$ is the conditional probability of transitioning from state $i$ to state $j$ provided the system starts in state $i$. That is, this condition ensures that the probability flow for all transitions between every pair of states is equivalent in both directions. Using our notation: 

\begin{equation}\label{eq4}
    (\pi_t)_i (m^K)_{ij} = (\pi_t)_j (m^K)_{ji}
\end{equation}

Since both $\pi_t$ and $m^K$ are sampled from the data, there is some uncertainty associated with both sides of equation \ref{eq4}. Therefore, the Detailed Balance Condition needs to be fulfilled within the confidence intervals associated with both sides of the equation \C{(see Materials and Methods for details and the project's repository for the raw numerical results on these values and confidence intervals)}. We find that equation \ref{eq4} is satisfied for almost every transition between edge states within the confidence intervals. \C{We observe small violations of the Detailed Balance Condition in some transitions between edge states. A table summarizing the characteristics of those transitions exhibiting a violation is provided in the Supplementary Material. Overall, the proportion of transitions that do not fulfil the condition is 0.003, 0.01, 0.04, 0.01, 0.023, 0.027, 0.004, 0.01, and 0.007 for transitions 1–9, respectively. In all cases, the magnitude of the violation is minor. Specifically, for the instances where a violation occurs, we have computed the z-scores, defined as the number of standard errors by which the condition is not met. In the majority of cases, the z-score is between 2 and 3, indicating that although the condition is not strictly fulfilled, it is very nearly so. Overall, given the low frequency and small magnitude of these violations, we conclude that they have a negligible effect on the} equilibrium of the network. It is possible to detect in Figure \ref{fig3} the effect this violation has in the equilibrium abundances. For the edges and triangles affected, the equilibrium curves are slightly above the empirical abundances, although there is still an almost perfect overlap between both. The method's ability to detect such a small violation supports the robustness of the equilibrium conditions found for the remaining transitions. \C{The raw results for the assessment of the detailed balance condition for degrees a triangles are included in the project's repository as well.}

\C{These observed violations might be attributed to simple statistical fluctuations in the network dynamics. Nonetheless, there appears to be a bias in the probability flows involved in these discrepancies favouring the formation of new positive relationships. Although these fluctuations remain compatible with the equilibrium framework we present, it is worthwhile to explore the potential origins of these discrepancies. One plausible explanation is that these deviations may be linked to the fact that we analyse students during a critical developmental stage. During this period, individuals are likely enhancing their cognitive capacities and their ability to manage a larger number of social connections. Being this indeed the case, the equilibrium dynamics framework may be applicable only to short- and medium-term dynamics during these life stages. However, it is important to stress that this interpretation is speculative. The overall results align very closely with the network being in equilibrium, suggesting that, regardless of the underlying cause, these discrepancies have only a minimal effect on the network dynamics.}\\

\C{As a final complementary analysis, we examined whether a metric capturing more complex structural behaviour—specifically, Betweenness Centrality—exhibits a pattern similar to that observed for degrees, edges, and triangles. Remarkably, as we can see in Figure \ref{fig4}, we observe that, in this case as well, the transition matrices are stationary, the detailed balance condition is fulfilled, and the empirical abundances coincide with the theoretical equilibrium distributions (see the  Supplementary Material for all technical details). All raw numerical results are available in the project's repository.}

\section*{Discussion}

In this paper, we have shown that the dynamical process driving the evolution of the personal relationships network of students belonging to a high school in Madrid, from 2020 to 2024, is stationary, and that the network itself is in dynamical equilibrium. We find that the abundances of edges, triangles, degrees \C{and even a more complex metric like the Betweenness Centrality,} are statistically equivalent in all snapshots and can be accurately predicted by the average transition matrix. Furthermore, a key feature of equilibrium in the statistical mechanics sense, Detailed Balance, is also verified with very small, point deviations. 

To the best of our knowledge, this result goes much beyond what has been studied so far. Some previous research reported a certain stability in the data analysed \cite{Katz1959,Hallinan1978,Doreian1996,Kucharski2018,Kossinets2006,Saramaki2014,Lubbers2010,Escribano2023}, but they neither explored this stability further nor showed rigorously that an empirical network is in equilibrium. In addition, as we mentioned in the introduction, other studies have found stability and robustness in social behaviour in broader populations. For instance, in \cite{Hobbs2017}, they show that social networks exhibit a resilience mechanism after the death of a friend, recovering the same number of active connections over time through increased interactions between friends of the decedent. This would be an example of how the network evolves to maintain its structural integrity. They also show that interactions between friends stabilize after a year following the death of a friend, similar to the stationarity in the transition probabilities in our findings. Moreover, they mention the possibility of the existence of a lower bound on individuals’ level of social connection, that would force them to replace lost friendships more quickly than they are driven to establish friendships in general. Within the context of mobility patterns, in \cite{Alessandretti2018}, the authors found that the number of locations an individual visits regularly is conserved over time, even while individual routines are unstable in the long term because of the continuous exploration of new locations. Furthermore, they find a connection between this number and the Dunbar's number, establishing a relation between these mobility patterns and the maintenance of their social relations. In \cite{Crectu2022,Iniguez2023,Avalle2024} the authors explore how online social behaviour follows consistent patterns over time, with \cite{Iniguez2023} focusing on the role played by ties of different strengths, and proving how different dynamical mechanisms compete for this stability to arise, an idea we will develop on the following discussion. Finally, in \cite{Iacopini2024} the authors show how different group social behaviours are consistent within both students and adult groups, pointing towards the generality in the dynamical behaviour of social bonds.

While the specificities of our sample population may limit direct extrapolation to other social networks, we have further reasons to believe that this result generalizes to broader populations beyond the alignment with the aforementioned studies. As discussed in the introduction, the temporal evolution of any network of personal relationships is influenced mainly by three important factors, operating locally in the network through what we call mechanisms. Thus, the first and second factor (attributes of people, endogenous structural mechanisms) operate together to control how and why these networks evolve at the level of links. As reviewed in the introduction, there is a spectrum of mechanisms through which these two factors affect the network structure. To name a few: the reciprocation of friendships, the creation of homophilic relationships, the establishment of transitive and hierarchical structures, the formation of balanced triangles, the avoidance of conflicts or the resource allocation of individuals. This last mechanism will be specially relevant, how people allocate their limited cognitive and material resources in their relationships to cover their social necessities. Regarding the third factor (institutional context of the network), the institutional setting of the society shapes the contexts in which relationships are formed, affecting who interacts with whom. In a high school, for instance, the fact that students share the same course, classes, or subjects biases the network structure. 

We now argue that, within this theoretical framework, the concept of equilibrium arises naturally. Basically, in the evolution of personal relationships, the context, the institutional setting, all influence with whom we interact (and, hopefully, with whom we establish relationships). Once this context is stable, people establish relationships allocating their cognitive and material resources among the people with whom they share this social environment. In the broader picture of the network, the other dynamical mechanisms shape how these relationships are structured (in simple terms, a person allocates some resources to establish their relationships, and dynamical mechanisms determine who these people are within the broader network). Thus, every time a relationship disappears, there is a liberation of resources that can be reinvested in new relationships. This argument is supported by the findings we mentioned from \cite{Hobbs2017}. At the level of dynamical mechanisms, there is a balancing process as well. For instance, when this new relationship is established, the reciprocation mechanisms will force it to be reciprocal. Nonetheless, there are other mechanisms destroying this reciprocity, like the formation of hierarchical structures. Similarly, there is a tendency towards the creation of balanced triangles, but also a tendency to avoid conflict and a high random component in conflict, that promotes the destruction of these triangles and the creation of unbalanced ones. Therefore, there is a competition between mechanisms having opposing effects on the network structures. The equilibrium is the expected outcome of the network dynamics. \C{It is worth mentioning that modelling the evolution of the network by integrating these spectrum of mechanisms is a task we are addressing in a future study, trying to explicitly tests all hypotheses presented here.}

An important insight from our results is the role the third factor plays in the observation of the equilibrium (the context, the institutional setting). Basically, to observe the network in equilibrium one needs to analyse the system at a proper temporal and spatial scale. The equilibrium would appear only after the network has had time to stabilize after the transient time, and the context of the observed network needs to be wide enough to be able to potentially saturate the social relationships of the people in the network. Let us illustrate this last point with an example. If we observe only a certain class within the school, it may happen that some people cease being friends within the class and compensate these relationships with people in other classes. However, if we are observing only this specific class, we may see how some relationships simply disappear from the system under our observation. If we draw an analogy with a physical system, to see the equilibrium it is necessary to observe the system within some `natural boundaries' that contain all the relevant dynamics of the system. Thus, in a social system, to detect the dynamical equilibrium, it is necessary to observe the network within some `natural boundaries', and these boundaries are defined by the context, social foci and organizational structure that is able to contain the majority of social relationships of the group. If not, you may observe certain stability in the system, but you may not be able to detect the equilibrium properly. In our case, the school seems to provide a perfect environment for this. Students may have other relationships outside the school, like familiar ties and contacts from extra-curricular activities, but the majority of their relationships are contained within the school, specially in these ages. Of course, the social foci can change in larger time scales. The organizational structure can slowly change, and in the case of a high school, at some point students leave, disrupting their network of relationships \cite{MayaJariego2022}. Nonetheless, within the school, as a closed environment that saturates the possible relationships students can have, and observing the system at a proper temporal scale in which the context remains stable, this equilibrium exists as a consequence of the competition and balance between mechanisms, and between necessities and resources allocation. 

\C{Although we propose that our findings may generalize to broader populations, we do not claim that every specific result presented here will replicate exactly in other contexts. We acknowledge that our study is based on a specific sample—a high school network representing a particular demographic group within a distinct cultural and institutional setting. Our main argument is that equilibrium behaviour emerges from a balance between competing mechanisms and between necessities and resources allocation. However, the precise operation of these mechanisms may vary across different sample groups. In practice, this implies that one might observe different equilibrium distributions, with variations in the abundances of degrees, edges, and triangles, and differences in the transition matrices. For example, as adolescents mature, they may develop improved emotional regulation and conflict management skills, resulting in fewer conflicts and reduced volatility. Nonetheless, we believe that even in such settings, the system would still reach an equilibrium state driven by the interplay of these mechanisms, albeit in a context-dependent manner.}

We want to stress that in our dataset only 15\% of nodes are present both in the first and the last waves. Thus, the equilibrium does not arise just because some fixed group of people coexist for a certain period of time. This result supports our view of the equilibrium as a consequence of the competition of dynamical mechanisms and a trade-off between resources allocation and social necessities more than a property of individuals themselves. Furthermore, this result indicates that statistically, people leaving and entering the school are similar, as the observed structure remains stable even when most of the network composition changes. This finding supports the view that there is not large variability in the structures of people's individual relationships, as highlighted by \cite{GonzalezCasado2024}, and our conclusions' applicability to many other social networks.

\C{It is important to recognize that in the work presented here we assume that the system is Markovian, meaning that each subsequent network snapshot depends solely on the current snapshot, with no influence from past states. In other words, we assume that the system has no memory. Nonetheless, we acknowledge that the network snapshots are likely not entirely independent, and that previous positive and negative links may influence future connections. For example, a new positive relationship might emerge with a higher probability between two individuals who previously shared a positive connection, or with a lower probability between those who had a conflict. While our results demonstrate remarkable accuracy in predicting the stationary state under the Markovian assumption—and we do not expect memory effects to significantly alter the overall distributions and aggregated transition matrices—we believe that such effects may still impact the network's evolution. Specifically, memory effects could modify the probability of certain transitions based on past interactions, so that even if the aggregated probabilities remain unchanged, individual edges might exhibit more variability in their activity over time. This would lead to a heterogeneity in transition probabilities that is not captured by the Markovian approximation. We recognize that this topic deserves further investigation, and we plan to develop a more comprehensive analysis of memory effects and non-Markovian dynamics in future work.}

In any case, if indeed social networks are generally in equilibrium, the findings of studies using cross-sectional data could potentially be generalized to the entire unobserved evolution of the network. When studied at the proper temporal and spatial scales after the transient, the observation of single snapshots can provide robust information about the dynamical mechanisms driving the network evolution and the structural features observed. This provides solid grounds for using methods like Exponential Random Graph Models and other methods that compare the network observation to null-models to understand the structural trends observed. From the social sciences perspective, this finding can avoid the difficulties in collecting longitudinal data on social networks, reducing the burden on respondents. Also, it opens the door for the design of intervention studies. Since the network properties are stable over time, the researcher can introduce interventions and associate the observed changes to the intervention isolated from other potential drivers of the network evolution.  \C{For example, consider an integration intervention aimed at promoting the inclusion of individuals at the lower end of the positive in-degree distribution, enhancing their visibility and likeability within the network. The impact of this intervention could be evaluated by comparing the resulting degree distributions with the previously observed equilibrium distributions. Similarly, an intervention focused on improving emotional regulation and relationship management might promote more balanced interactions within the network. By analyzing the edge-type distributions before and after such an intervention, one could determine whether the frequency of imbalanced relationships (e.g., +2+1, +1+0, and +2+0) decreases while balanced positive interactions (e.g., +1+1 and +2+2) become more prevalent.}
 
Finally, from a theoretical modelling perspective, this equilibrium has direct implications for the predictability of social behaviour and our understanding of the system's interdependence. The resulting observed social networks are the product of competing mechanisms and resources management, and while relationships are dynamic and constantly evolving, influenced by cognitive, emotional and environmental factors, this evolution is highly constrained, and there is a tendency toward stable structures. The development of simple mathematical models to describe human behaviour is justified by these insights. For instance, our result supports the application of statistical mechanical methods to social systems. Statistical mechanics often reveals universal patterns in physical systems, suggesting that human relationships may also exhibit this type of universal behaviours. This conclusion connects with the findings of \cite{Forecasting2023}, that deals with the predictability of societal changes, stressing the importance of combining such statistical models with domain-specific knowledge to make better predictions. Such an observation challenges the view that social systems are uniquely complex and unpredictable, raising the question of whether social phenomena can be explained by general laws or if they are inherently unique and context-dependent. To properly answer this question, future longitudinal studies should be conducted to establish whether this observed equilibrium is indeed a general property of social networks.

\section{Materials and methods}
\subsection{Materials}
In this section we provide details about the data collection process, the data composition and the data curation.
\subsubsection*{Data collection}
The collection of our data was performed through surveys administered in the school via a computer interface. To elicit relationships, students were presented with a list of all other students in the high school. They were then asked to select individuals with whom they had a relationship. Specifically, the questionnaire included the following question: `You can now see the list of all the students in the school. Please mark those you have any relationship with by clicking `very good relationship', `good relationship', `bad relationship' or `very bad relationship'. Only one choice is possible. If you do not mark any option, it will be understood to mean that you do not have a relationship with the person'. Typically, it took students about 15 minutes to complete the survey, and they were supervised by a school teacher throughout the process. Our study was approved by the Institutional Review Board of the UC3M, which stipulated an opt-out procedure. We should note that there were no opt-outs, effectively eliminating any potential selection bias. The only students who did not participate were those who were absent on the day of the experiment.

\subsubsection*{Data composition and data curation} 

In Table \ref{table1}, we present the composition of the network snapshot by snapshot. The missing people column includes people who were absent on the day of the survey because there were no opt-outs. We removed some outliers from the analyses, defined as people with more than 30 outgoing very good relationships, more than 50 outgoing good relationships, more than 15 outgoing bad relationships or more than 15 outgoing very bad relationships. These numbers were selected by comparing outgoing with incoming degree distributions. The proportion of outliers removed is shown in the Outliers column. In any case, results with and without outliers are approximately equal, showing that their presence would not change our conclusions.

\begin{table}[t!]
\centering
\caption{Composition of each network snapshot. In the columns Sex, Missing Data and Outliers, numbers represent proportions.}
\label{table1}
\begin{tabular}{lcccc}
Wave & Respondents & Sex (M/F) & Missing Data & Outliers \\
\midrule
DEC 2020 & 409 & 0.52/0.48 & 0.11 & 0.03 \\
MAY 2021 & 409 & 0.52/0.48 & 0.10 & 0.03 \\
SEP 2021 & 530 & 0.49/0.51 & 0.06 & 0.04 \\
FEB 2022 & 530 & 0.49/0.51 & 0.09 & 0.04 \\
MAY 2022 & 530 & 0.49/0.51 & 0.00 & 0.07 \\
SEP 2022 & 524 & 0.54/0.46 & 0.06 & 0.03 \\
JAN 2023 & 535 & 0.54/0.46 & 0.09 & 0.07 \\
MAY 2023 & 536 & 0.54/0.46 & 0.13 & 0.05 \\
SEP 2023 & 554 & 0.51/0.49 & 0.06 & 0.04 \\
JAN 2024 & 563 & 0.51/0.49 & 0.07 & 0.05 \\
\bottomrule
\end{tabular}
\end{table}

\subsubsection*{Data availability}
All the aggregated data necessary to replicate our results can be found in \href{https://github.com/miguelangel-gonzalezc/equilibrium_in_social_networks}{this repository}. \C{Besides, in the same repository one can find all the codes necessary to reproduce our computations.}  

\subsection{Methods}

\subsubsection*{Missing Data Treatment}
\C{In reconstructing the network snapshots from the survey data, a link from node \textit{i} to node \textit{j} can be absent for two distinct reasons. First, it may indicate that person \textit{i} deliberately chose not to declare any relationship—positive or negative—with person \textit{j}. Alternatively, the absence of a link may result from person \textit{i} being absent on the day of the survey, so that the link is missing simply because no response was recorded. In our analysis, we have carefully distinguished between these two cases. Throughout the paper, we use `+0' to denote an absent link that results from a deliberate decision by the respondent not to assign any weight. This is distinct from `missing links', which refer to potential links that are absent due to non-response. Missing links are excluded from all counts to avoid introducing bias. For example, a notation such as `+0+1' indicates that the first person deliberately chose not to reciprocate the link, whereas if the data were missing, that particular link would not be counted in the frequency statistics. All missing data are systematically removed from our computations (including those for stationarity matrices, detailed balance, abundance measures, etc.). For instance, if there is a missing link from \textit{i} to \textit{j} and a +1 link from \textit{j} to \textit{i}—which would form an edge noted as (miss)+1—this case is not counted as a +0+1 edge. Similarly, if in the next wave both individuals are present and we observe a +1+1 link, this is not regarded as a transition from a +0+1 to a +1+1 state.}

\subsubsection*{Construction of the transition matrix}

Let us introduce the concept of a transition matrix using the edges as our macroscopic property. In the case of edges, there are 25 possible edge states. If we take two consecutive snapshots of the network, for each individual edge, we can record what kind of transition it went through. A +2+2 edge can stay a +2+2, or become a +2+1, or even a +2-1, etc. If we repeat this process for all the links, we can construct a 25x25 matrix in which each row is the edge state before the transition, in the first snapshot, and each column is the edge state after the transition, in the second snapshot. The $ij$ element of this matrix would be the number of edges starting in state $i$ and ending in state $j$. In this matrix, we can divide each element by the total count in each row. By doing this, the $ij$ element in this final matrix represents $P(j|i)$, i. e., the conditional probability of an edge of going to the $j$ state provided it started in the $i$ state. We call this matrix the transition matrix (also known as the Markov Matrix).

\subsubsection*{Statistical comparison of transition matrices}

\begin{figure}[t!]
\centering
\includegraphics[width=0.8\linewidth]{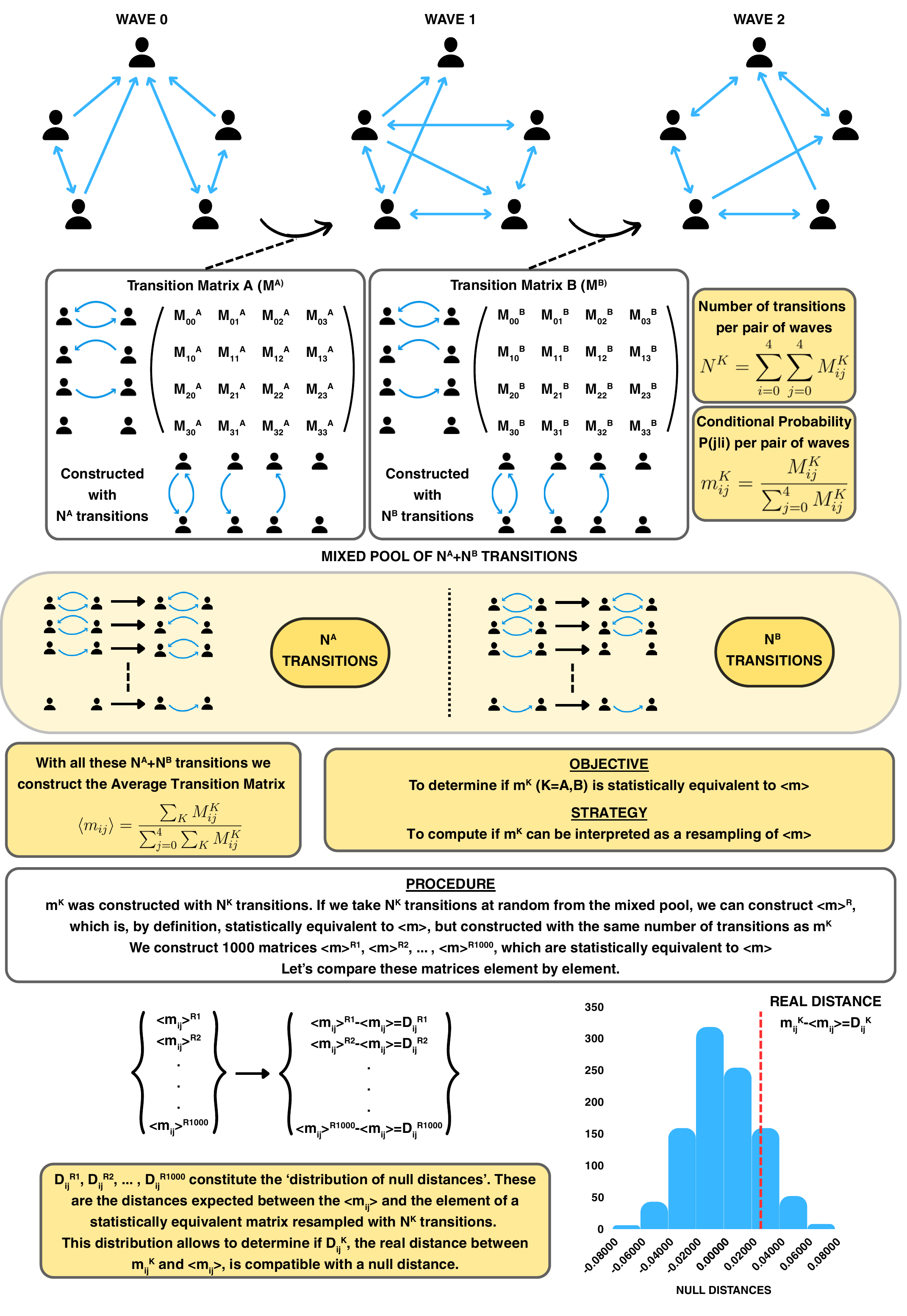}
\caption{\footnotesize Illustration of the method used to construct and compare the transition matrices. To simplify the explanation, an example with three snapshots of a directed network of fixed composition is shown, where links take a value of either 0 or 1.}
\label{fig5}
\end{figure}

\C{The purpose is to analyse whether the nine transition matrices are statistically equivalent to one another. However, to simplify the process, we instead check whether each of the nine transition matrices is statistically equivalent to the average transition matrix computed from all the transitions observed between consecutive snapshots. Let us explain the method to perform this statistical comparison. Besides, in Figure \ref{fig5} we have included a diagram with the main points of the procedure. }

\C{Let us introduce some notation using the dynamics of edges as an example. To construct each transition matrix, we record all transitions between edge states that occur from one snapshot to the next. Denote a transition by the letter \(K = A, B, C, \dots\), and define \(N^K\) as the number of transitions observed between one snapshot and the next, for that particular case. For instance, \(N^A\) is the number of transitions recorded from snapshot 0 to snapshot 1, which we use to construct the transition matrix \(M^A\) (containing the transition counts as defined in the previous section) and \(m^A\) (containing the corresponding conditional probabilities, i.e., \(P(j|i)\), as defined previously). In summary, \(M_{ij}^A\) represents the number of transitions observed from edge type \(i\) to edge type \(j\) between snapshot 0 and snapshot 1, and \(m_{ij}^A\) represents the conditional probability \(P(j|i)\) for that interval. We construct one such matrix for each pair of consecutive snapshots.}

\C{By aggregating the observed transitions from all pairs of consecutive snapshots, we form a pool of transitions. This pool contains \(N^A\) transitions from the first pair, \(N^B\) transitions from the second pair, and so on, so that the total number of transitions is \( N = N^A + N^B + N^C + \cdots\) By aggregating all these transitions, we can construct the Average Transition Matrix \(\langle m \rangle\) for the entire observed process.}

\C{With this setup in mind, our objective is to determine whether \(m^K\) (for \(K = A, B, C, \dots\)) is equivalent to \(\langle m \rangle\); that is, whether each individual transition matrix is statistically equivalent to the average matrix. The key idea is that if the process is stationary, each transition matrix between consecutive snapshots is simply a resampling of the average matrix because the underlying process remains constant over time. In other words, we ask whether \(m^K\) can be interpreted as a resampling of \(\langle m \rangle\).}

\C{The procedure is as follows. The matrix \(m^K\), which we wish to compare with \(\langle m \rangle\), is constructed from the \(N^K\) observed transitions between consecutive snapshots. To assess its equivalence to \(\langle m \rangle\), we compare it with resampled versions of the average matrix. Specifically, by randomly selecting \(N^K\) transitions from the pool of all observed transitions, we effectively construct a transition matrix that is, by definition, statistically equivalent to \(\langle m \rangle\), but computed with the same number of transitions as \(m^K\). We generate 1000 such matrices, denoted \(\langle m \rangle^{R1}\), \(\langle m \rangle^{R2}\), \(\dots\), \(\langle m \rangle^{R1000}\), to represent random samples statistically equivalent to the average matrix. Our goal is to determine whether these resampled matrices behave similarly to \(m^K\).}

\C{For each element \(ij\) of the matrices, we compute the distance between the resampled matrices and the average matrix:
\[
D_{ij}^{Rx} = \langle m_{ij} \rangle^{Rx} - \langle m_{ij} \rangle \quad (x = 1,2,\dots,1000).
\]
These distances form what we call the `distribution of null distances', which represents the expected differences between \(\langle m_{ij} \rangle\) and the corresponding element of a statistically equivalent matrix resampled with \(N^K\) transitions.}

\C{Next, we compare the observed distance
\[
D_{ij}^{K} = m_{ij}^{K} - \langle m_{ij} \rangle,
\]
with this null distribution. If \(D_{ij}^K\) is compatible with the distribution of null distances, then for element \(ij\) the real transition matrix \(m^K\) behaves like a resample of \(\langle m \rangle\), and they can be considered statistically equivalent. Note that for each element \(ij\) we obtain a distribution of null distances against which the observed distance is compared.}

\C{From the distribution of null distances, we compute the following quantities for each element \(ij\) of each transition matrix:
\begin{itemize}
    \item \textbf{p-value}: Computed as the proportion of times the absolute observed distance is smaller than the absolute null distance, i.e., 
    \[
    \Pr\Bigl(|D_{ij}^K| < |D_{ij}^{Rx}|\Bigr).
    \]
    \item \textbf{z-score}: Computed as 
    \[
    z = \frac{D_{ij}^K - \mu(D_{ij}^{Rx})}{\sigma(D_{ij}^{Rx})},
    \]
    where \(\mu(D_{ij}^{Rx})\) and \(\sigma(D_{ij}^{Rx})\) are the mean and standard deviation of the null distance distribution, respectively. This z-score indicates how many standard deviations the observed distance deviates from the mean of the null distribution.
\end{itemize}
In principle, choosing a significance level of 0.05, we would consider \(m_{ij}^K\) statistically equivalent to \(\langle m_{ij} \rangle\) if the p-value exceeds 0.05. However, because this method involves multiple comparisons across the elements of the matrix, we apply a Bonferroni correction and consider the two elements statistically equivalent only if the p-value exceeds (0.05/number of comparisons). For elements with p-values below this threshold, we analyze the computed z-scores to assess how far the observed element \(ij\) deviates from the null distribution.}

\subsubsection*{Estimation of confidence intervals}
\C{In our analysis, whenever we need to estimate the confidence interval associated with a probability or a density, computed as \(p=x/M\), where \(x\) is the number of cases that belong to the category of interest and \(M\) is the total number of cases, we use the approximation to the variance of the Binomial distribution such that a 0.95 confidence interval is given by:
\[
CI(0.95) \approx p \pm 1.96 \sqrt{\frac{p(1-p)}{M}}.
\]
All confidence intervals presented in the paper correspond to 95\% confidence intervals.}

\section{Acknowledgements}
M.A.G.-C. acknowledges support from the Comunidad de Madrid through the grants for the hiring of pre-doctoral research personnel in training (reference PIPF-2023/COM-29487). M.A.G.-C. and A.S. acknowledge support from grant PID2022-141802NB-I00 (BASIC) funded by MCIN/AEI/10.13039/501100011033 and by ‘ERDFA way of making Europe’, and from grant MapCDPerNets---Programa Fundamentos de la Fundaci\'on BBVA 2022. A. S. T acknowledges support by FCT – Fundação para a Ciência e Tecnologia – through the LASIGE Research Unit, ref. UID/000408/2025.

\section{Author Contributions}
All authors conceived and conceptualized the research, A.S. collected the data, M.A.G.-C. curated the data, formalized the analyses and obtained the results, and all authors discussed and interpreted the results and wrote the manuscript.

% \section{Significance Statement}
% We show that the network of personal relationships among 900 people over the course of four years is in dynamical equilibrium, meaning that the collective properties of the network remain stable while the microscopic dynamics of relationships are highly active, with individuals entering and leaving the network. This finding implies that all pictures of the network are statistically equivalent. Thus, a single snapshot is enough to understand its structure, serving as a justification for many traditional research methods. We argue this may be a general feature of social systems arising from competing mechanisms of relationship evolution. This result lays the foundation for understanding and predicting key features of social systems from a general viewpoint.

\bibliographystyle{unsrt}
\bibliography{references}  %%% Uncomment this line and comment out the ``thebibliography'' section below to use the external .bib file (using bibtex) .

\end{document}